\documentclass[pra,twocolumn,preprintnumbers,amsmath,amssymb,letterpaper]{revtex4}

\usepackage{graphicx}
\DeclareGraphicsExtensions{.eps}

\begin{document}

\title{Low-Temperature and High-Pressure Induced Swelling of a Hydrophobic 
       Polymer-Chain in Aqueous Solution}

\author{Dietmar Paschek} 
\email{dietmar.paschek@udo.edu}
\author{Sascha Nonn}
\author{Alfons Geiger}
\affiliation{Physikalische Chemie, Universit\"at Dortmund, 
  Otto-Hahn-Str. 6, D-44221 Dortmund, Germany}


\date{\today}

\begin{abstract}
We report molecular dynamics simulations of a hydrophobic polymer-chain
in aqueous solution between $260\,\mbox{K}$ and $420\,\mbox{K}$
at pressures of $1\,\mbox{bar}$, $3000\,\mbox{bar}$, and $4500\,\mbox{bar}$.
The simulations reveal a hydrophobically collapsed state at low pressures
and high temperatures. At $3000\,\mbox{bar}$ and about $260\,\mbox{K}$
and at $4500\,\mbox{bar}$ and about $260\,\mbox{K}$, however, a
transition to a swelled state is observed. The transition is driven by
a smaller volume and a remarkably strong lower enthalpy of the swelled
state, indicating a steep positive slope of the 
corresponding transition line. The swelling is stabilized almost completely
by the energetically favorable state of water in the polymers
hydrophobic first hydration shell at low temperatures. 
Although surprising, this finding is consistent with the observation
of a positive heat capacity of hydrophobic solvation.
Moreover, the slope and location of the
observed swelling transition for the collapsed hydrophobic chain coincides 
remarkably well with the cold denaturation transition of proteins.
\end{abstract}

\maketitle

\section{Introduction}

\label{sec:intro}

Hydrophobic effects have been shown to be of relevance for a wide range
of physicochemical and biophysical phenomena
\cite{Tanford,Ben-Naim:Hydrophobic,Blokzijl:93}. 
In particular, they 
are seen as an important driving force with regard to
the folding of
proteins \cite{Privalov:88,Southall:2002,Pratt:2002}. Consequently, a wealth of studies
of hydrophobic interactions 
using molecular simulation techniques
have been undertaken over the past three decades
\cite{Geiger:79,Pangali:79,Zichi:85,Smith.D:92,Smith.D:93,Pearlman:93,Forsman:94,Skipper:96,Luedemann:96,Luedemann:97,Ghosh:2001,Ghosh:2002,Ghosh:2003,Shimizu:2000,Shimizu:2001,Shimizu:2002,Rick:97,Rick:2000,Rick:2003,Paschek:2004:1,Paschek:2004:2}. 
Simulation studies have revealed that the contact state of a pair of hydrophobic
particles in aqueous solution is entropically
stabilized at ambient conditions \cite{Smith.D:92,Smith.D:93}. 
In addition, the contribution of the solvation 
heat capacity has been recognized recently
\cite{Shimizu:2000,Shimizu:2001,Shimizu:2002,Rick:97,Rick:2000,Rick:2003,Paschek:2004:1,Paschek:2004:2}. 
In general, the dissolution of hydrophobic particles is accompanied with an increase of
the associated heat capacity \cite{Wilhelm:77,Rettich:81,Naghibi:86}. As a
consequence, the dissolution of a hydrophobic particle is found to be increasingly 
enthalpically stabilized with decreasing temperature \cite{Paschek:2004:2}.

Biopolymers such as proteins remain stable and functional only in a limited 
pressure and temperature range \cite{Privalov:1990,Royer:2002,Smeller:2002}. 
Increasing temperature lead to structural changes differing from the native folded state. 
This is often accompanied with large fluctuations and 
aggregation phenomena. 
Hence, pressure effects on proteins are of 
interest in biotechnology and biology \cite{Ludwig:1999},
as pressure is shifting the equilibrium of protein
configurations without increasing thermal fluctuations 
\cite{Zipp:1973,Hawley:71,Brandts:1970,Silva:2001}.
Proteins undergo unfolding upon addition of pressures above
$2\,\mbox{kbars}$.  High pressures are also 
able to dissociate protein complexes \cite{Silva:1992, Peng:1993}. 
The solvent water plays a crucial role in the effect
of pressure in protein unfolding  \cite{Hawley:71,Smeller:2002,Panick:1999,Herberhold:2002}
and the addition of co-solvents is found to have a significant influence on
the size, location and shape of the stability region of proteins \cite{Herberhold:2004,Ravindra:2004}.

At high pressures ($>\!2\,\mbox{kbars}$), the volume of proteins upon unfolding
decreases. This seemed to be inconsistent with the assumption that protein
unfolding is equivalent to the transfer of hydrophobic groups from the protein
interior to the aqueous solvent, since the volume change upon transfer of
hydrophobic groups to water are positive. 
Hummer et al.\ \cite{Hummer:98:1} suggested a scenario
in which pressure unfolding of proteins is modeled as the transfer of water
into the protein hydrophobic core with increasing pressure. The transfer of
water molecules into the protein interior is essential for the pressure unfolding
process, leading to the dissociation of close hydrophobic contacts and
subsequent swelling of the hydrophobic protein interior through insertions of
water molecules \cite{Hummer:98:1}. The characteristic features of water-mediated
interactions between hydrophobic solutes in water are found to be
pressure-dependent. In particular, with increasing pressure the
solvent-separated configurations in the solute-solute potential of mean force
is  stabilized with respect to the contact configurations. In addition,
the desolvation barrier increases monotonically with respect to both contact
and solvent-separated configurations. The locations of the minima and the
barrier move toward shorter separations, and pressure effects are considerably
amplified for larger hydrophobic solutes \cite{Ghosh:2001,Ghosh:2002,Ghosh:2003}.

Pressure also changes the entropy/enthalpy balance of the hydrophobic
interactions.  Ghosh et al. found that the contact minimum is dominated by entropy, whereas 
the solvent-separated minimum is stabilized by favorable enthalpy of
association \cite{Ghosh:2001,Ghosh:2002}. 
Both the entropy and enthalpy at the contact minimum seem to change
little with increasing pressure leading to the relative pressure insensitivity of the
contact minimum configurations. In contrast, the solvent-separated
configurations are increasingly 
stabilized at higher pressures by enthalpic contributions that prevail over the 
slightly unfavorable entropic contributions to the free energy \cite{Ghosh:2002}. 

In this contribution, we focus particularly on the scenario 
proposed by Hummer et al.\ \cite{Hummer:98:1}
of water penetrating into the protein interior at elevated pressures. 
We study, however, a very much simplified model system of protein: 
A fully hydrated polymer-chain, consisting of 20 interconnected
hydrophobic particles. The polymer-chain approach
has been recently advocated by Chandler and co-workers
\cite{Huang:2000,Rein_ten_Wolde:2002,Chandler:2005}, suggesting that the 
collapse of a hydrophobic polymer chain is driven by a drying transition.
In Ref. \cite{Rein_ten_Wolde:2002} the hydrophobic chain is
modeled by repulsive polymer/water interactions only, whereas the solvent is
represented by a coarse grained model.
In the recent work of Ghosh et al. \cite{Ghosh:2005}, molecular
dynamics simulations of a polymer-chain 
in an explicit solvent
reveal the effect of salt concentrations on
the polymer configuration, increasingly favoring compact folded
configurations of the polymer with increasing salt concentration.
In the above mentioned studies 
a ``stiff'' polymer chain was employed, which exhibits a stretched
equilibrium configuration in absence of a solvent. Thus the
energy needed to bend the polymer-chain is used to counterbalance the
tendency to minimize the solvent accessible surface.
In the present study no such terms are employed, hence 
a chain of linked hydrophobic particles, free of
bond-angles and dihedral potential barriers is considered.
Thus the reported
configurational changes of the polymer 
have to be completely attributed to the
influence of the solvent. 

\section{Computational Methods}

\subsection{MD Simulation details}

\label{sec:MD}

We report molecular dynamics (MD) simulations of a purely hydrophobic 
polymer-chain dissolved in an aqueous solution.
The polymer-chain consists of 20 hydrophobic polymer
beads, represented by Lennard-Jones interaction sites with 
$\sigma_{XX}\!=\!3.975\,\mbox{\AA}$, 
$\epsilon_{XX}\,k_B^{-1}\!=\!214.7\,\mbox{K}$ \cite{Paschek:2004:1}.
The water-polymer cross parameters were obtained using the conventional
Lorentz-Berthelot mixing rules
with $\sigma_{ij}\!=\!\left(\sigma_{ii}+\sigma_{jj}\right)/2$ and
$\epsilon_{ij}\!=\!\sqrt{\epsilon_{ii}\epsilon_{jj}}$. The polymer sites
are linked by rigid bonds of $4.2\,\mbox{\AA}$ length. This bond-length was
determined to represent the bead-bead contact distance for non-linked
particles in aqueous solution \cite{Paschek:2004:2}. In addition, this value
corresponds roughly to the distance between adjacent hydrophobic Valine sidechains in
a polypeptide. All intramolecular non-bonded interactions were taken into account, 
except interactions between adjacent bonded sites.
No additional bond-bending or torsional potentials were used.
The water phase is represented by 1000 TIP5P water molecules \cite{Mahoney:2000}.
The TIP5P model was chosen,
since it represents the hydrophobic solvation behavior of water 
on the low-temperature side probably most realistically among the
simple point charge models \cite{Paschek:2004:1}.
Moreover, the temperature dependent strength of the hydrophobic interaction 
was found to be quite critically linked to the temperature dependence of
waters expansivity \cite{Paschek:2004:1}. The simulations discussed here
 were carried over a broad temperature range
at pressures of $1\;\mbox{bar}$, $3000\;\mbox{bar}$,
 and $4500\;\mbox{bar}$. Individual 
MD-simulations extend up to $100\,\mbox{ns}$, while the total simulation time
adds up to about $0.76\,\mu\mbox{s}$. For completeness, a detailed simulation
protocol is given in Table \ref{tab:protocol}.

The MD-simulations are carried out
in the NPT ensemble using
the Nos\'e-Hoover thermostat 
\cite{Nose:84,Hoover:85}
and the Rahman-Parrinello barostat 
\cite{Parrinello:81,Nose:83} with
coupling times $\tau_T\!=\!1.5\,\mbox{ps}$ and 
$\tau_p\!=\!2.5\,\mbox{ps}$
(assuming an isothermal compressibility of
$\chi_T\!=\!4.5\;10^{-5}\,\mbox{bar}^{-1}$), respectively.
The electrostatic interactions are treated
in the ``full potential'' approach
by the smooth particle mesh Ewald summation 
\cite{Essmann:95} with a real space
cutoff of $0.9\,\mbox{nm}$ and a mesh spacing of approximately
$0.12\,\mbox{nm}$ and 4th order
interpolation. The Ewald convergence factor $\alpha$ was set to
$3.38\,\mbox{nm}^{-1}$ (corresponding to a relative accuracy of
the Ewald sum of $10^{-5}$).
A $2.0\,\mbox{fs}$ 
timestep was used for all simulations. Solvent constraints were solved
using the SETTLE procedure \cite{Miyamoto:92},
while the SHAKE-algorithm was used for the polymer constraints \cite{Ryckaert:77}.
For all simulations reported here 
the GROMACS 3.2  program \cite{gmxpaper,gmx32} was used.
Statistical errors in the analysis
were computed using the method of Flyvbjerg and Petersen \cite{Flyvbjerg:89}.
For each system an initial equilibration run of about 
$1\,\mbox{ns}$ length was performed using the Berendsen 
weak coupling scheme for pressure and temperature control
($\tau_T\!=\!\tau_p\!=\!0.5\,\mbox{ps}$) \cite{Berendsen:84}. 
\begin{table}[!t]
  \centering
  \renewcommand{\tabcolsep}{0.38cm}
  \renewcommand{\arraystretch}{1.0}
  \small
  \begin{tabular}{cccc} \hline\hline \\[-4pt]
  $T/\mbox{K}$ &
  $\tau/\mbox{ns}$&
  $\left< V\right>/\mbox{nm}^3$  & 
  $\left< E\right>/\mbox{kJ}\,\mbox{mol}^{-1}$ 
\\[6pt] \hline \\[-6pt]
\multicolumn{4}{c}{$1\,\mbox{bar}$:}\\[0.2em]
$260$ & 80   &  $31.803\pm0.006$   &  $-38065\pm6$ \\
$280$ & 54   &  $31.710\pm0.002$   &  $-35036\pm5$  \\
$300$ & 30   &  $32.008\pm0.003$   &  $-32469\pm6$  \\
$320$ & 30   &  $32.563\pm0.003$   &  $-30128\pm4$  \\
$340$ & 30   &  $33.324\pm0.003$   &  $-27936\pm5$  \\
$380$ & 27   &  $35.450\pm0.020$   &  $-23799\pm20$  \\
$420$ & 28   &  $38.880\pm0.050$   &  $-19772\pm30$  \\[0.6em]
\multicolumn{4}{c}{$3000\,\mbox{bar}$:}\\[0.2em]
$260$ & 100  &  $27.868\pm0.003$   &  $-37432\pm9$ \\
$280$ & 100  &  $28.043\pm0.002$   &  $-35138\pm3$  \\
$300$ & 50   &  $28.305\pm0.002$   &  $-33067\pm3$  \\
$360$ & 30   &  $29.428\pm0.002$   &  $-27535\pm2$  \\ 
$400$ & 30   &  $30.384\pm0.001$   &  $-24252\pm3$  \\[0.6em]
\multicolumn{4}{c}{$4500\,\mbox{bar}$:}\\[0.2em]
$280$ & 73   &  $26.935\pm0.002$   &  $-35222\pm3$  \\
$300$ & 73   &  $27.212\pm0.001$   &  $-33220\pm3$  \\
$320$ & 25   &  $27.514\pm0.002$   &  $-31360\pm3$  \\
$360$ & 25   &  $28.206\pm0.001$   &  $-27918\pm3$  
\\[6pt] \hline\hline
  \end{tabular}
  \caption{\footnotesize Simulation protocol for the 
    performed MD-simulations.
    $\tau$: Simulation time; $\left< V\right>$: Average box-volume; 
    $\left< E\right>$: Average total energy.}
  \label{tab:protocol}
\end{table}

It should be mentioned that the completely stretched polymer extends
to about $8\,\mbox{nm}$, which exceeds the used box length about 2.5 times.
However, in practice, contacts between the polymer and its virtual image were
not observed during the simulation runs discussed here.

\subsection{Energy Analysis}

\label{sec:energy}

In order  to assign potential energies to
individual molecules and thus to be able to distinguish
between contributions from the ``hydration shell'' and 
from the ``bulk'',
we determine energies by
a reaction field method based on the minimum image cube.
This ``cubic'' cutoff procedure has been
originally proposed by Neumann \cite{Neumann:83}.
Roberts and Schnitker \cite{Roberts:94,Roberts:95} 
have shown that
the obtained energy estimates are comparing very
well with the Ewald-summation including tin-foil boundary conditions.

For convenience, we divide the potential energies in contributions
assigned to the individual molecules with
\begin{eqnarray}
  E
  &=&
  \sum_{i=1}^M E_i \nonumber \\
  E_i
  &=&
  \left(\frac{1}{2}\sum_{j=1}^M E_{ij}\right) + E_{i,corr.} \;,
\end{eqnarray}
where $E_i$ is the potential energy assigned to molecule $i$,
$M$ is the total number of molecules. The molecule-molecule
pair energy
\begin{eqnarray}
  E_{ij}
  &=&
  \sum_\alpha \sum_\beta 
  4\,\epsilon_{i\alpha j\beta}
  \left[
    \left(
      \frac{\sigma_{i\alpha j\beta}}{r_{i\alpha j\beta}}
    \right)^{12}
    -
    \left(
      \frac{\sigma_{i\alpha j\beta}}{r_{i\alpha j\beta}}
    \right)^{6}
  \right]
  \nonumber\\
 & &
  \hspace*{3em}+ \frac{q_{i\alpha}q_{j\beta}}{r_{i\alpha j\beta}}
\end{eqnarray}
is then obtained as sum over discrete interaction sites $\alpha$
and $\beta$, with $r_{i\alpha j\beta}\!=\!|\vec{r}_{j\beta}-\vec{r}_{i\alpha}|$
based on the molecule/molecule center of mass minimum image separation.
We employ long range corrections
$E_{i,corr.}\!=\!E^{el}_{i,corr.}+E^{LJ}_{i,corr.}$
accounting for electrostatic, as well as Lennard
Jones interactions. The electrostatic correction 
\begin{eqnarray}
  E^{el}_{i,corr.}
  &=&
  \frac{2\pi}{3 V}\vec{D}\,\vec{d}_i
\end{eqnarray} 
is a reaction field term, corresponding to the
cubic cutoff, assuming an infinitely large 
dielectric constant. Here 
$\vec{d}_i\!=\!\sum_\alpha q_{i\alpha}\vec{r}_{i\alpha}$
is the dipole moment of
molecule $i$, $\vec{D}=\sum_i \vec{d}_i$ 
is the total dipole moment of all molecules in the simulation cell
and $V$ is the instantaneous volume of the simulation box.
Finally, also the long range Lennard-Jones corrections for 
to  the minimum image were taken into
account, as outlined in Ref. \cite{Paschek:2004:2}.

\section{Results}

\label{sec:results}

\begin{figure}[!b]
  \centering
  \includegraphics[angle=0,width=8cm]{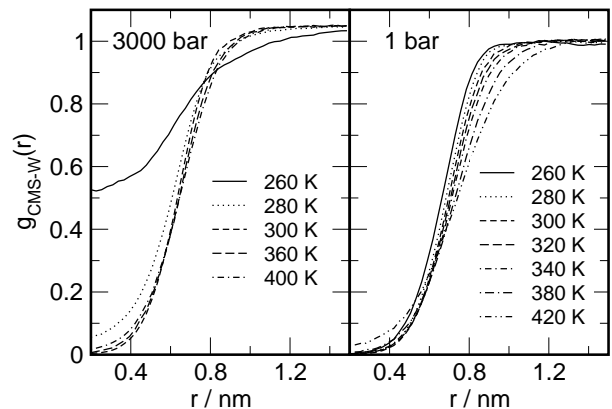}

 \caption{\small Polymer/water center of mass pair correlation functions
   for all simulated state points.
  }
  \label{fig:04}
\end{figure}

\begin{table*}[!t]
  \centering
  \renewcommand{\tabcolsep}{0.7cm}
  \renewcommand{\arraystretch}{1.0}
  \small
  \begin{tabular}{lcccc} \hline\hline \\[-4pt]
 \multicolumn{1}{c}{Bulk:} &
 \multicolumn{2}{c}{$260\,\mbox{K}$; $3000\,\mbox{bar}$} &
 \multicolumn{2}{c}{$280\,\mbox{K}$; $4500\,\mbox{bar}$} \\[0.2em]
$E(\mbox{H}_2\mbox{O})/\mbox{kJ}\,\mbox{mol}^{-1}$ &
 \multicolumn{2}{c}{$-43.657\pm0.004$} &
 \multicolumn{2}{c}{$-42.081\pm0.003$} \\
$V_m(\mbox{H}_2\mbox{O})/\mbox{cm}^3\,\mbox{mol}^{-1}$ &
 \multicolumn{2}{c}{$16.02\pm0.05$} &
 \multicolumn{2}{c}{$15.50\pm0.02$} 
\\[0.6em]\hline\\[-0.6em]
   \multicolumn{1}{c}{Hydration Shell:} &
 \multicolumn{2}{c}{$260\,\mbox{K}$; $3000\,\mbox{bar}$} &
 \multicolumn{2}{c}{$280\,\mbox{K}$; $4500\,\mbox{bar}$} \\[0.2em]
      &
  ``collapsed''&
  ``swelled'' &
  ``collapsed''&
  ``swelled'' \\[2pt]
$N(\mbox{H}_2\mbox{O})$     
&  $126.5\pm2.1$     &  $184.2\pm2.8$ 
&  $129.4\pm1.3$     &  $173.0\pm2.0$ \\
$V(\mbox{Shell})/\mbox{nm}^3$     
&  $4.51\pm0.09$     &  $5.96\pm0.08$  
&  $4.49\pm0.03$     &  $5.58\pm0.07$  \\
$V_m(\mbox{H}_2\mbox{O})/\mbox{cm}^3\,\mbox{mol}^{-1}$ 
& $21.47\pm0.05$ &  $19.49\pm0.05$  
& $20.97\pm0.07$ &  $19.46\pm0.07$  \\
$E(\mbox{H}_2\mbox{O})/\mbox{kJ}\,\mbox{mol}^{-1}$
&  $-43.63\pm0.05$   &  $-44.43\pm0.03$ 
&  $-41.85\pm0.02$   &  $-42.34\pm0.03$ \\
$E(\mbox{Polymer})/\mbox{kJ}\,\mbox{mol}^{-1}$
&  $-149.7\pm0.5$   &  $-141.3\pm0.6$ 
&  $-141.2\pm0.2$   &  $-126.7\pm0.5$ 
\\[6pt] \hline\hline
  \end{tabular}
  \caption{\footnotesize Energies and volumes obtained for the bulk and
    hydration shell for the given statepoints.
    The ``hydration shell'' is defined as the 
    volume with a distance $R\!\leq\!0.52\,\mbox{nm}$ to any polymer site,
    whereas the ``bulk'' is obtained for distances of $R\!>\!1.0\,\mbox{nm}$.
    $V(\mbox{Shell})$ denotes the corresponding volume  occupied by the polymer and its
    first hydration shell. $N(\mbox{H}_2\mbox{O})$ gives the average number of
    hydration shell water molecules. The hydration shell water molar
    volumes are according to 
    $V_m(\mbox{H}_2\mbox{O})\!=\!$V(\mbox{Shell})/$N(\mbox{H}_2\mbox{O})*N_A$,
    where $N_A$ is the Avogadro number.
    The potential energies $E$ of the polymer and water in bulk and shell 
    are given per molecule. ``collapsed'' and ``swelled'' states are defined
    by $R_G\!<\!0.65\,\mbox{nm}$ and $R_G\!>\!0.8\,\mbox{nm}$, respectively.
  }
  \label{tab:hshell}
\end{table*}
\begin{figure}[!b]
  \centering
  \includegraphics[angle=0,width=5cm]{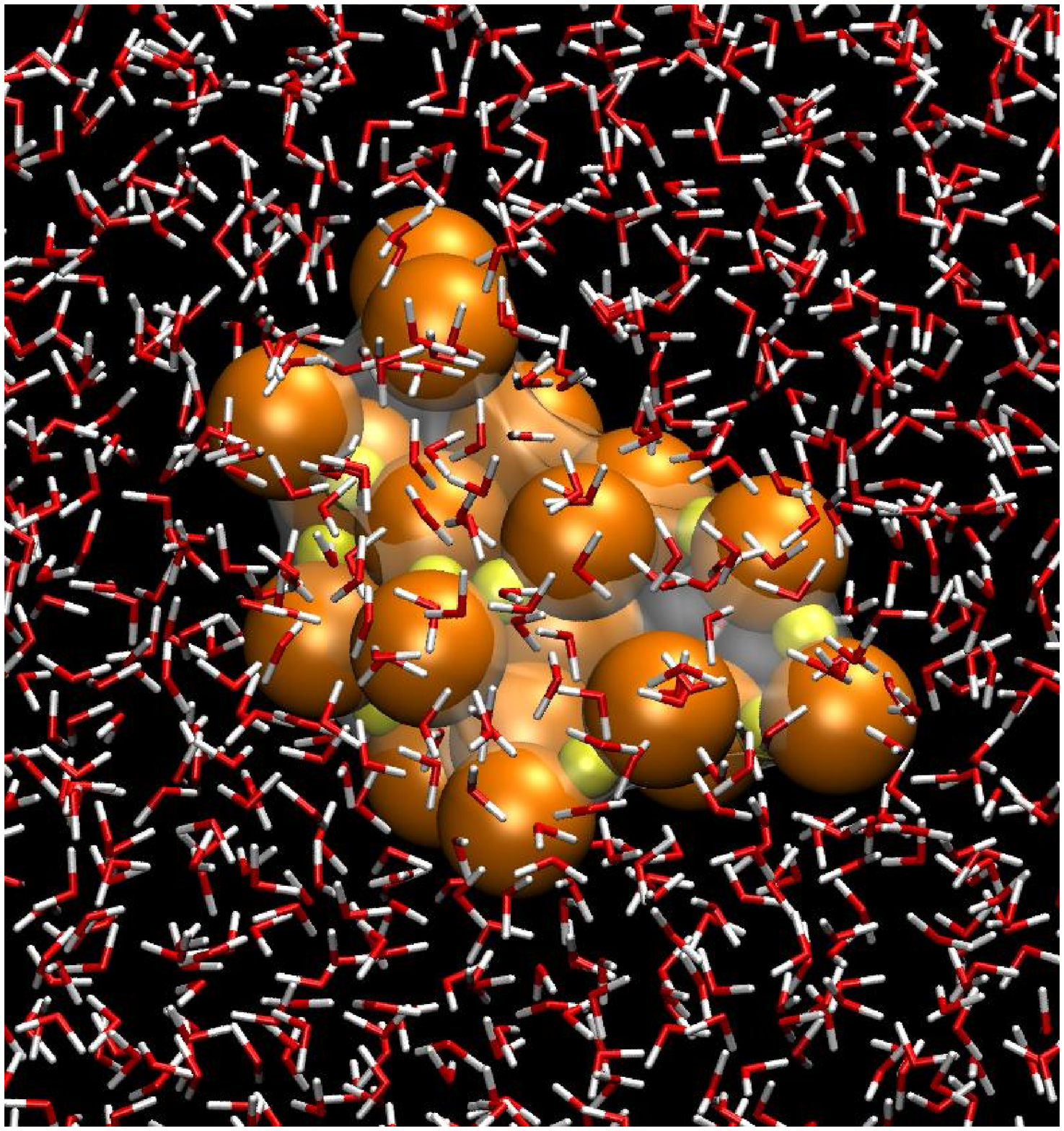}

  \vspace{0.3cm}
  \includegraphics[angle=0,width=5cm]{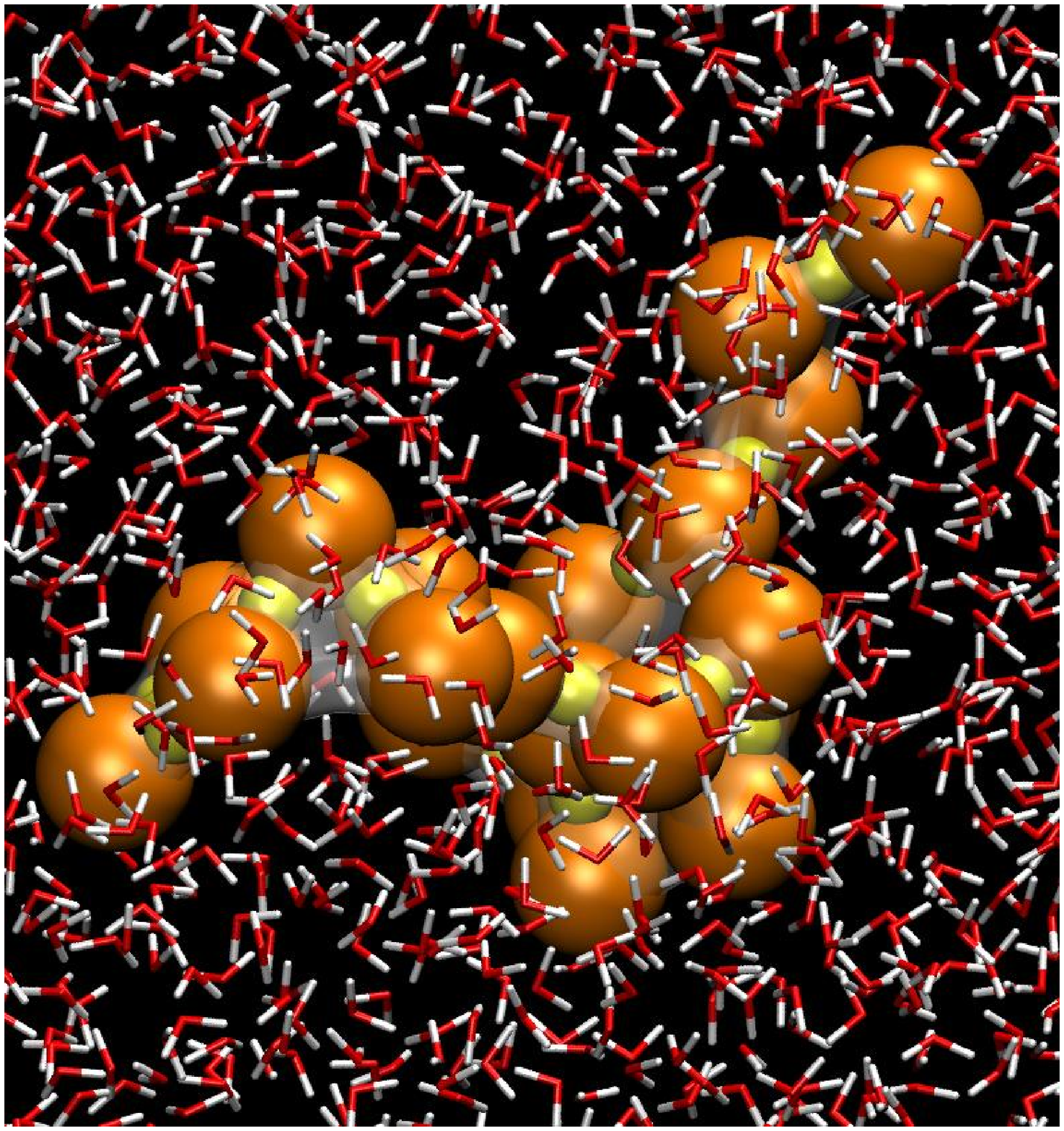}

  \caption{\small Representative configurations of the hydrophobic 
    polymer in aqueous solution as obtained from simulations
    $3000\;\mbox{bar}$. Top: Collapsed configuration observed
    at $300\;\mbox{K}$. Bottom: Swelled configuration observed at
     $260\;\mbox{K}$.
  }
  \label{fig:01}
\end{figure}
\begin{figure*}[!t]
  \centering
  \includegraphics[angle=0,width=12cm]{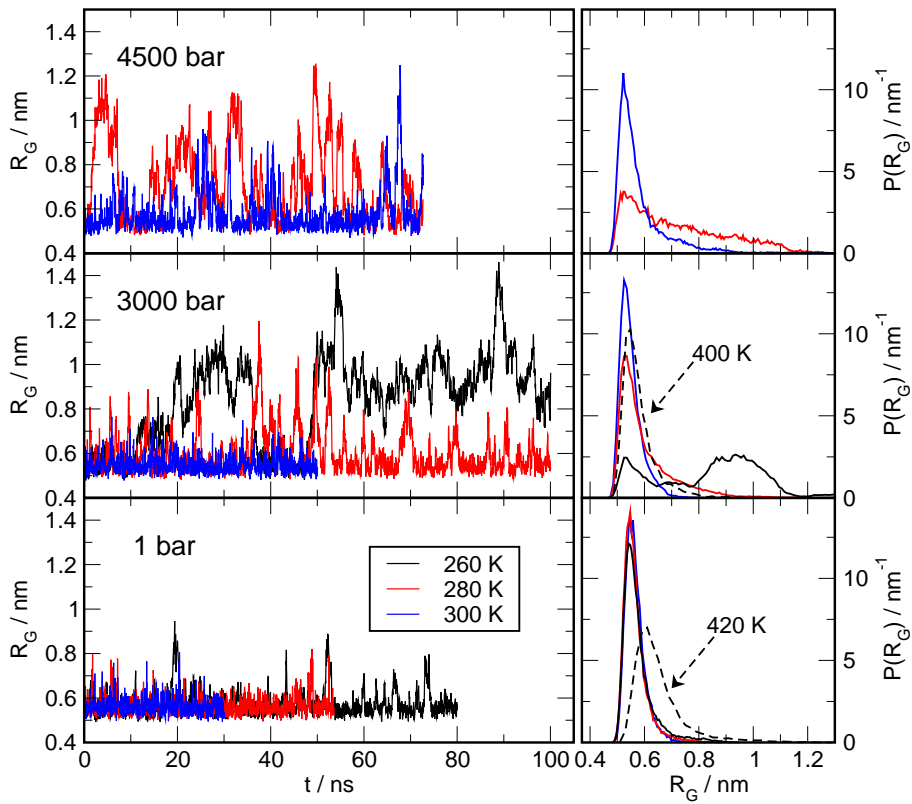}

  \caption{\small 
    Time-evolution and corresponding probability density distribution
    of the polymers radius of gyration.
    Shown are the lowest temperatures at $1\,\mbox{bar}$ (bottom), 
    $3000\,\mbox{bar}$  and
    $4500\,\mbox{bar}$ (top). The dashed line indicates the probability
    distribution obtained for the highest temperatures, respectively.
  }
  \label{fig:02}
\end{figure*}
\begin{figure}[!b]
  \centering
  \includegraphics[angle=0,width=6.5cm]{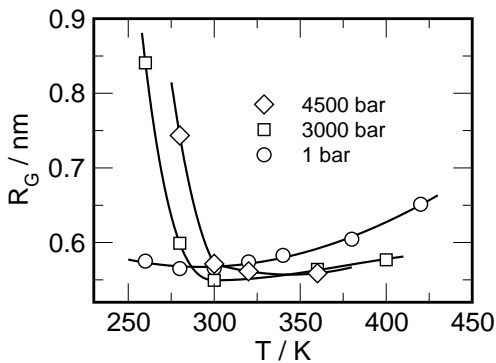}

  \caption{\small Average radius of gyration $R_G$ of the hydrated
    polymer as obtained for all temperatures at
    $1\,\mbox{bar}$, $3000\,\mbox{bar}$ and $4500\,\mbox{bar}$.
    The lines are drawn to guide the eye.
  }
  \label{fig:03}
\end{figure}
The structure of the dissolved hydrophobic polymer-chain is characterized by
its radius of gyration 
$R_G^2\!=\!1/20\sum_{i=1}^{20} \left(\vec{r}_i-\vec{c} \right)^2$
with  $\vec{c}=1/20\sum_{i=1}^{20} \vec{r}_i$.
The typical conformation of this polymer at high temperatures
and low pressures is
a compact, collapsed state with 
a $R_G$ between $0.5\,\mbox{nm}$ and $0.6\,\mbox{nm}$,
as shown in Figure \ref{fig:03}. A snapshot of a representative collapsed-chain
configuration is shown in Figure \ref{fig:01}. 
The conformational distribution with respect to $R_G$ is found
to be narrow, with a half width at half maximum of $R_G$ of 
about $0.1\,\mbox{nm}$. Test-simulations at
$300\,\mbox{K}$ and ambient pressure conditions, starting
with a swelled configuration of $R_G\!\approx\!1\,\mbox{nm}$, show 
a collapse on a timescale $<\!1\,\mbox{ns}$.
The polymer/water center of mass pair correlation functions
(given in Figure \ref{fig:04}) reveal that in the collapsed state
water is completely excluded from the polymer interior.

The temperature dependence of the average radius of gyration 
at a pressure of $1\,\mbox{bar}$ is characterized
by a shallow minimum at about $300\,\mbox{K}$. 
With increasing temperature the distribution of $R_G$
more or less maintains
its shape, but is becoming broader with its maximum shifting to
larger values as shown in Figure \ref{fig:02}. The high-temperature
behavior is more or less similar for all pressures discussed here.
In addition, a slight penetration of water into the polymer interior 
is observed for $420\mbox{K}$ (see Figure \ref{fig:04}). 

\begin{figure}[!t]
  \centering
  \includegraphics[angle=0,width=7.5cm]{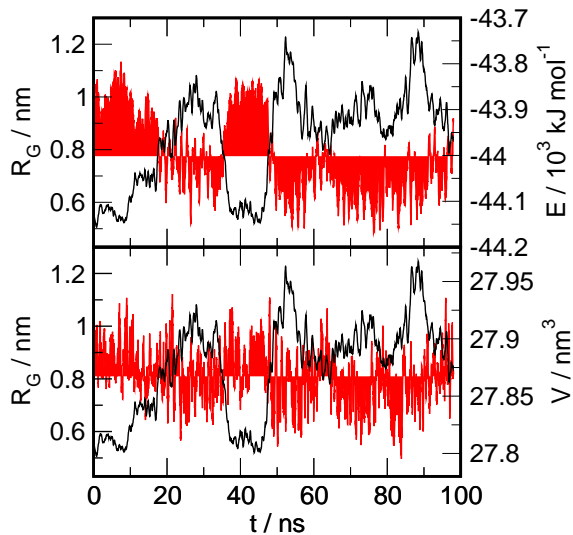}

 \caption{\small 
   Time evolution of the radius 
   of gyration of the polymer (black),
   potential energy (red, top) and volume (red, bottom)
   as obtained from the simulation at $260\,\mbox{K}$ and
   $3000\,\mbox{bar}$. 
   Potential energies and box-volumes were directly 
   taken from GROMACS simulation output. In order to reduce the
   noise, all data were smoothed using with the same
   Savitzky-Golay filter
   \cite{NumRecipes}.
  }
  \label{fig:05}
\end{figure}

At lower temperatures, however, a different behavior is starting to emerge. From the time evolution of $R_G$,
shown in Figure \ref{fig:02} it is evident that at about $260\,\mbox{K}$
and $1\,\mbox{bar}$ the polymer increasingly starts to explore extended-chain
configurations.
These extended-chain states are occurring only rather infrequently and are
short-lived with a life-time of about $1\,\mbox{ns}$.
We would like to emphasize that the $280\,\mbox{K}$/$3000\,\mbox{bar}$ 
and $300\,\mbox{K}$/$4500\,\mbox{bar}$  trajectories show a
similar behavior. Again, the extended-chain configurations are
rather short-lived with the chain quickly returning to the collapsed state.
In a quantitative manner, an increase of the  population of swelled
states is found, which is
due to the apparent higher frequency of large amplitude $R_G$-fluctuations.
The most striking difference, however, is observed
for $260\,\mbox{K}$ at $3000\,\mbox{bar}$  and
$280\,\mbox{K}$ at $4500\,\mbox{bar}$ .
Here the extended-chain configurations are dominating, although
an equilibrium between collapsed and swelled states is still maintained. 
The radius of gyration is showing an apparently bimodal distribution 
at $260\,\mbox{K}$ at $3000\,\mbox{bar}$ and a
broad distribution at  $280\,\mbox{K}$ at $4500\,\mbox{bar}$.
The representation of both (swelled and compact) at each of these
states, strongly suggests that the corresponding conformational transition temperatures 
are located in close proximity to the temperatures indicated.
The temperature
dependence of $R_G$ at different pressures, as given in Figure \ref{fig:03},
suggests that the high-temperature collapsed-chain state is more compact at
elevated pressures, which would correspond to a more compressed coiled state.
At $280\,\mbox{K}$ and $3000\,\mbox{bar}$ and  
$260\,\mbox{K}$ and $4500\,\mbox{bar}$, however, the situation has already changed, and the
tendency to explore more extended configurations leads to an increase in
$R_G$ which is quickly progressing upon cooling. 
The low temperature destabilization of the
collapsed state is in line with the decreased stability
of hydrophobic contacts observed for elevated pressures \cite{Hummer:98:1,Ghosh:2002,Ghosh:2003}. 
Figure \ref{fig:03} implies that the transition towards a swelled state at  $3000\,\mbox{bar}$
and  $4500\,\mbox{bar}$  occurs 
in a rather narrow temperature interval, 
suggesting a rather large enthalpy difference 
between collapsed and swelled state. In other words: 
it shows a large
``cooperativity''. Thermodynamical consistency requires that
the swelled low temperature state has to be energetically more stable
than the compact state and that it has to occupy a smaller total volume.
Figure \ref{fig:05} shows a superposition of the time-evolution 
of the potential energy and box-volume, as well as
$R_G$ for $260\,\mbox{K}$ at $3000\,\mbox{bar}$. In order to make the
trends more clearly visible, the noise in the fluctuations have been reduced by Savitzky
Golay filtering \cite{NumRecipes}. Both, the potential energy
as well as the box-volume are clearly anti-correlated with respect to $R_G$.
The swelled state of the polymer is enthalpically
stabilized with energy difference
of $\Delta E_u\!=\!E(\mbox{swell.})-E(\mbox{coll.})$ of 
$-182\,\mbox{kJ}\,\mbox{mol}^{-1}$ at $3000\,\mbox{bar}$
and of 
 $-82\;\mbox{kJ}\,\mbox{mol}^{-1}$ at $4500\,\mbox{bar}$.
The swelled state leads also to smaller box volumes, with
$\Delta V_u\!=\!V(\mbox{swell.})-V(\mbox{coll.})$ of 
$-20\;\mbox{ml}\,\mbox{mol}^{-1}$ at $3000\,\mbox{bar}$
and of 
$-10\,\mbox{ml}\,\mbox{mol}^{-1}$. 
We would like to point out the observed enthalpy and volume
changes are of similar magnitude as observed for some 
proteins \cite{Royer:2002} and peptides \cite{Nicolini:2004}. 
\begin{figure}[!t]
  \centering
  \includegraphics[angle=0,width=6.5cm]{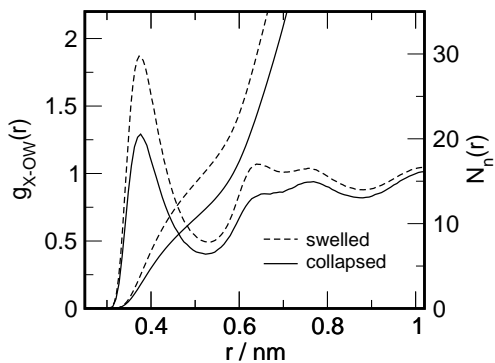}

  \caption{\small Polymer-bead/water-oxygen site-site pair correlation function
    $g_{\mbox{\footnotesize X}-\mbox{\footnotesize OW}}(r)$ 
    and integrated number of nearest neighbors $N_n(r)$
    for the collapsed and swelled states  at $260\,\mbox{K}$ and
    $3000\,\mbox{bar}$. 
  }
  \label{fig:06}
\end{figure}
\begin{figure}[!t]
  \centering
  \includegraphics[angle=0,width=8cm]{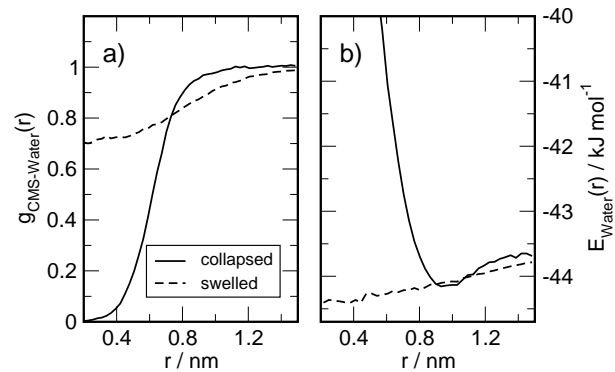}

  \caption{\small 
    (a) Polymer/water center of mass pair correlation function
    for the collapsed and swelled states  at $260\,\mbox{K}$ and
    $3000\,\mbox{bar}$.
    (b) Potential energy of the water molecules as a function
    of distance to the polymer center.
  }
  \label{fig:07}
\end{figure}
\begin{figure}[!t]
  \centering
  \includegraphics[angle=0,width=6.5cm]{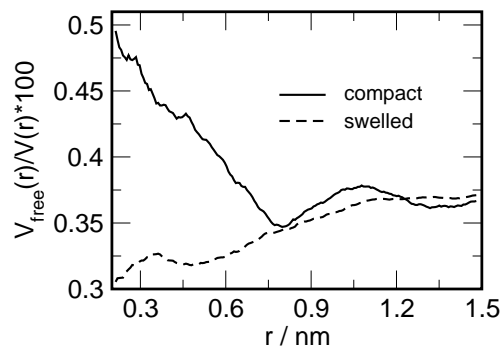}

  \caption{\small 
    Free volume fraction accessible to a
    small hard sphere particle as a function of
    distance to the polymer center. Given are dependencies
    for the collapsed and swelled states at $260\,\mbox{K}$ and
    $3000\,\mbox{bar}$. The diameter of the particle with
    $\sigma_{Y-OW}\!=\!2.5\,\mbox{\AA}$ and
    $\sigma_{Y-X}\!=\!2.93\,\mbox{\AA}$ 
    is scaling as the corresponding Lennard-Jones sigmas.
  }
  \label{fig:08}
\end{figure}

In the following paragraphs we would like to show that
the energy {\em and} volume change can be attributed almost quantitatively to 
the first hydration shell of the polymer and have to
be almost exclusively attributed to the solvent. 
Figure \ref{fig:06} depicts the polymer/water-oxygen site-site pair correlation functions
for the collapsed and swelled states
obtained at $260\,\mbox{K}$ and $3000\,\mbox{bar}$. The first minimum, located
at a distance of $0.52\,\mbox{nm}$, indicates the spatial dimension the
first hydration shell.
The average number of water neighbors around each polymer site changes about 
$40\,\%$ from
$11.5$ to $16$ upon swelling (at $3000\,\mbox{bar}$). 
At $4500\,\mbox{bar}$ a qualitatively similar behavior is observed.
In order to quantify the changes in
the hydration shell, we define the volume occupied by the
polymer and the first hydration shell as the volume with a distance of
$R\leq0.52\,\mbox{nm}$ with respect to any polymer-site. 
The properties obtained for the hydration shell of the collapsed and
swelled polymer and the water bulk
are summarized in Table \ref{tab:hshell}. 

The polymer/water center of mass pair correlation function as well as the potential
energy of water as a function distance to the polymer center for
the collapsed and swelled states are given in Figure \ref{fig:07}.
From Figure \ref{fig:07} it is evident that for the region of the polymer/water
interface ($r\!\approx\!0.9\,\mbox{nm}$) water is $0.3\,\mbox{kJ}\,\mbox{mol}^{-1}$ 
more stable as in the bulk. However, due to the lack of  water
neighbors, a further penetration of individual waters
into the collapsed hydrophobic coil is energetically 
unfavorable, as the steep increase in Figure  \ref{fig:07}b
indicates. When drawing the balance over all water molecules in the hydration
shell, energy gains and penalties almost cancel out completely and the average 
potential energy of a water molecule in the hydration shell of the collapsed
polymer is nearly identical to the value observed for the water bulk
at $3000\,\mbox{bar}$ and slightly more unstable at  $4500\,\mbox{bar}$
(see Table \ref{tab:hshell}).
For the case of the swelled polymer, the energy penalty is absent
and the water in the hydration shell gains
$-0.78\,\mbox{kJ}\,\mbox{mol}^{-1}$ potential energy per water molecule with
respect to the bulk on average ($-0.26\,\mbox{kJ}\,\mbox{mol}^{-1}$ at  $4500\,\mbox{bar}$).
The potential energy difference between collapsed and swelled state is 
about  $\Delta E_u\!\approx\!
[E(\mbox{shell,swell.})-E(\mbox{bulk})] \,N(\mbox{shell,swell.})
-[E(\mbox{shell,coll.})-E(\mbox{bulk})] \,N(\mbox{shell,coll.})+
E(\mbox{polymer,swell.})-E(\mbox{polymer,colll.})
\!=\!-138\,\mbox{kJ}\,\mbox{mol}^{-1}$
($60 \,\mbox{kJ}\,\mbox{mol}^{-1}$)
which already accounts largely 
for the observed total energy difference of about 
$-182\,\mbox{kJ}\,\mbox{mol}^{-1}$ ($-82\,\mbox{kJ}\,\mbox{mol}^{-1}$).
The potential energy of the polymer changes just by
$8.4\,\mbox{kJ}\,\mbox{mol}^{-1}$  ($14.5\,\mbox{kJ}\,\mbox{mol}^{-1}$), 
which is due to the fact that the loss of
intramolecular interactions when going from the collapsed to the swelled state
is almost completely compensated by polymer/solvent interactions
(The values for $4500\,\mbox{bar}$ are given in parentheses). 
Hence the the extended-chain configurations at low temperatures 
appear to be largely solvent-stabilized.

The volume change upon swelling
is $\Delta V_{m,u}\!\approx\!
[V_m(\mbox{shell,swell.})-V_m(\mbox{bulk})]\,N(\mbox{shell,swell.})
-[V_m(\mbox{shell,coll.})-V(\mbox{bulk})]\,N(\mbox{shell,coll.})
\!=\!-57\,\mbox{ml}\,\mbox{mol}^{-1}$ 
at $4500\,\mbox{bar}$ and 
$\Delta V_{m,u}\!\approx\!-21\,\mbox{ml}\,\mbox{mol}^{-1}$
for the $4500\,\mbox{bar}$-isobar.
origin of the negative volume change upon swelling is indicated
in Figure \ref{fig:04}. Here the free volume fraction accessible
to a small sphere is shown as a function of distance to the polymer center. 
For the collapsed state, apparently
an excess free volume in the polymer interior 
is available to a hard sphere particle which is absent in
the swelled state. The hydration shell around the
hydrophobic chain appears to be more tightly packed, so that the average free free
volume fraction is even lower than for the water bulk.
Hence the increase in solvent accessible 
surface (increasing number of hydration shell waters) is
overcompensated by the decrease in molar volume of water in the swelled
hydration shell state. The hydration shell volume of the collapsed state includes, of
course, the volume of the hydrophobic core, which is made accessible to the solvent upon
unfolding of the chain. 

\begin{figure}[!t]
  \centering
  \includegraphics[angle=0,width=7cm]{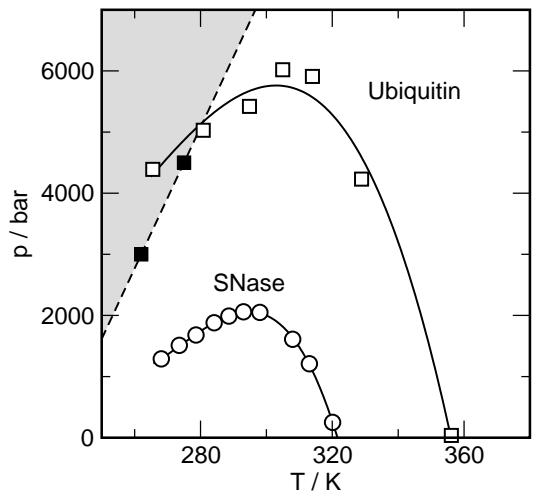}

  \caption{\small 
    Sketch of the swelling line of the
    hydrophobic polymer used in this study compared with
    experimental stability diagrams as obtained
    for SNase \cite{Panick:1999} and Ubiquitin \cite{Herberhold:2002}.
  }
  \label{fig:09}
\end{figure}

Finally, we would like to compare the behavior of the hydrophobic polymer
to the experimentally obtained stability diagrams of proteins.
Interpolating the $R_G$-data from Figure \ref{fig:03} and assuming
that the transition appears at $R_G\!\approx\!0.7\,\mbox{nm}$,
we obtain $T_u\!\approx\!268\,\mbox{K}$ and $282\,\mbox{K}$ for
 $3000\,\mbox{bar}$ and  $4500\,\mbox{bar}$, respectively.
Using the enthalpies $\Delta H_u\!\approx\!\Delta E_u+P\Delta V_u$ and
volumes $\Delta V_u$ according the data from Table \ref{tab:hshell},
we obtain a slope of the coexistence line of
a$d p_{eq}/dT\!=\!\Delta H_U/T_u\Delta
V_U$ of about $100\;\mbox{bar}\,\mbox{K}^{-1}$ 
for both pressures, seemingly consistent with the approximate transition
temperatures.  Although having more data
would be desirable, we might conjecture that the slope does not seem to change
much, but the enthalpy and volume differences tend to decrease with
increasing pressure. Since the differences might disappear completely 
at higher pressures, the swelling transition of the hydrophobic polymer 
probably ends up in a ``critical point''  (The term ``critical point'' might not
be fully appropriate here since swelling transition has a monotonous character
due the finite size of the polymer).
In Figure \ref{fig:09} the swelling
transition line, as well as the
stability diagrams for Staphyloccal Nuclease \cite{Panick:1999}
and ubiquitin \cite{Herberhold:2002} are shown.
We would like to point out that the location and slope of the
swelling transition shows remarkable similarity to the given
cold-denaturation lines.
The hydrophobic polymer seemingly behaves as suggested by
the water penetration scenario 
according Hummer et al.\ \cite{Hummer:98:1}. In the present
case the energy stabilization
of the swelled configuration is dominated the energy gain of the
hydration water. In a real polypeptide this is not necessarily true. 
Backbone hydration and the equilibrium between intra- and intermolecular
hydrogen bonds will certainly play an important role. How this delicate
balance might influence the equilibrium between swelled and collapsed
configuration should be further investigated. The high temperature
side of the protein stability diagram might not be accessible by
our simple model since it is largely related to internal secondary
structural transformation to a ``molten globule'' state 
\cite{Daggett:1992}.
In addition, we would like to mention that
the observed cold swelling of the hydrophobic polymer
has some similarity to the scenario proposed for the 
swelling of polymeric tropo-elastin upon cooling, as
advocated by the work of D.W.\ Urry \cite{Urry:93}.

In the present simulation study the energy gain upon unfolding is apparently a 
consequence of the increasing water-water pair interactions in the 
hydration shell at low temperatures \cite{Paschek:2004:2}.
The hydrogen bond network in the hydration shell starts disintegrating upon
heating more strongly as in the bulk 
since no hydrogen bonds can be formed to the hydrophobic Lennard-Jones
particles \cite{Paschek:2004:2}. 
The counterbalance between strengthened
hydrogen bonds and enhanced disintegration of the hydrogen bond network is
widely regarded as the mechanism leading to the positive solvation heat
capacity associated with hydrophobic hydration \cite{Silverstein:2000,Southall:2002}.
As a consequence, the observed transition temperature 
might be critically influenced by the change of the heat capacity of the solvent
in the hydration shell compared to the bulk. This might also shed light on the
mechanism how co-solvents affect the (cold) unfolding transition of proteins.

\section{Conclusion}

Molecular dynamics simulations of a hydrophobic polymer-chain
in aqueous solution between $260\,\mbox{K}$ and $420\,\mbox{K}$
at pressures of $1\,\mbox{bar}$, $3000\,\mbox{bar}$,
 and $4500\,\mbox{bar}$ reveal a hydrophobically collapsed state at low pressures
and high temperatures. At about $268\,\mbox{K}$ and $3000\,\mbox{bar}$ 
and at $282\,\mbox{K}$ and $4500\,\mbox{bar}$ a
transition to a swelled state is observed. The transition is driven by
a smaller volume and a remarkably strong lower enthalpy of the swelled
state. The volume effect is basically due to a smaller net-volume of the
extended hydrated state compared to the collapsed state exhibiting
hydrophobic cavity volumes and penetration of  
the internal volume by water. Moreover, the extended-chain 
structure is almost completely energetically stabilized by
the lower potential energy of the water molecules in the hydration shell.
Consequently, the increasingly stably water-water hydrogen bonds  close
to a hydrophobic particle, leading
the positive heat capacity of solvation, which is a 
signature for hydrophobic hydration \cite{Silverstein:2000}, 
is the key to the observed behavior.
The strong energy and volume differences 
indicates a steep positive slope of  the 
corresponding transition line of about $100\,\mbox{bar}\,\mbox{K}^{-1}$. 
The observed stability line for
the collapsed hydrophobic chain shows strong similarity with the 
lower temperature side of the stability digram of proteins in aqueous solution.

{\bf Acknowledgment.} Financial support by the Deutsche Forschungsgemeinschaft
(DFG-Forschergruppe 436) and 
University of Dortmund (Forschungsband ``Molekulare Aspekte der Biowissenschaften'')
is gratefully acknowledged. We would like to thank Angel E. Garc\'ia for
helpful comments.


\begin{thebibliography}{69}
\expandafter\ifx\csname natexlab\endcsname\relax\def\natexlab#1{#1}\fi
\expandafter\ifx\csname bibnamefont\endcsname\relax
  \def\bibnamefont#1{#1}\fi
\expandafter\ifx\csname bibfnamefont\endcsname\relax
  \def\bibfnamefont#1{#1}\fi
\expandafter\ifx\csname citenamefont\endcsname\relax
  \def\citenamefont#1{#1}\fi
\expandafter\ifx\csname url\endcsname\relax
  \def\url#1{\texttt{#1}}\fi
\expandafter\ifx\csname urlprefix\endcsname\relax\def\urlprefix{URL }\fi
\providecommand{\bibinfo}[2]{#2}
\providecommand{\eprint}[2][]{\url{#2}}

\bibitem[{\citenamefont{Tanford}(1980)}]{Tanford}
\bibinfo{author}{\bibfnamefont{C.}~\bibnamefont{Tanford}},
  \emph{\bibinfo{title}{The Hydrophobic Effect: Formation of Micelles and
  Biological Membranes}} (\bibinfo{publisher}{John Wiley \& Sons},
  \bibinfo{address}{New York}, \bibinfo{year}{1980}), \bibinfo{edition}{2nd}
  ed.

\bibitem[{\citenamefont{Ben-Naim}(1980)}]{Ben-Naim:Hydrophobic}
\bibinfo{author}{\bibfnamefont{A.}~\bibnamefont{Ben-Naim}},
  \emph{\bibinfo{title}{Hydrophobic Interactions}} (\bibinfo{publisher}{Plenum
  Press}, \bibinfo{address}{New York}, \bibinfo{year}{1980}).

\bibitem[{\citenamefont{Blokzijl and Engberts}(1993)}]{Blokzijl:93}
\bibinfo{author}{\bibfnamefont{W.}~\bibnamefont{Blokzijl}} \bibnamefont{and}
  \bibinfo{author}{\bibfnamefont{J.~B. F.~N.} \bibnamefont{Engberts}},
  \bibinfo{journal}{Angew. Chem.} \textbf{\bibinfo{volume}{105}},
  \bibinfo{pages}{1610} (\bibinfo{year}{1993}).

\bibitem[{\citenamefont{Privalov and Gill}(1988)}]{Privalov:88}
\bibinfo{author}{\bibfnamefont{P.~L.} \bibnamefont{Privalov}} \bibnamefont{and}
  \bibinfo{author}{\bibfnamefont{S.~J.} \bibnamefont{Gill}},
  \bibinfo{journal}{Adv. Protein Chem.} \textbf{\bibinfo{volume}{39}},
  \bibinfo{pages}{191} (\bibinfo{year}{1988}).

\bibitem[{\citenamefont{Southall et~al.}(2002)\citenamefont{Southall, Dill, and
  Haymet}}]{Southall:2002}
\bibinfo{author}{\bibfnamefont{N.~T.} \bibnamefont{Southall}},
  \bibinfo{author}{\bibfnamefont{K.~A.} \bibnamefont{Dill}}, \bibnamefont{and}
  \bibinfo{author}{\bibfnamefont{A.~D.~J.} \bibnamefont{Haymet}},
  \bibinfo{journal}{J. Phys. Chem. B} \textbf{\bibinfo{volume}{106}},
  \bibinfo{pages}{521} (\bibinfo{year}{2002}).

\bibitem[{\citenamefont{Pratt and Pohorille}(2002)}]{Pratt:2002}
\bibinfo{author}{\bibfnamefont{L.~R.} \bibnamefont{Pratt}} \bibnamefont{and}
  \bibinfo{author}{\bibfnamefont{A.}~\bibnamefont{Pohorille}},
  \bibinfo{journal}{Chem. Rev.} \textbf{\bibinfo{volume}{102}},
  \bibinfo{pages}{2671} (\bibinfo{year}{2002}).

\bibitem[{\citenamefont{Geiger et~al.}(1979)\citenamefont{Geiger, Rahman, and
  Stillinger}}]{Geiger:79}
\bibinfo{author}{\bibfnamefont{A.}~\bibnamefont{Geiger}},
  \bibinfo{author}{\bibfnamefont{A.}~\bibnamefont{Rahman}}, \bibnamefont{and}
  \bibinfo{author}{\bibfnamefont{F.~H.} \bibnamefont{Stillinger}},
  \bibinfo{journal}{J. Chem. Phys} \textbf{\bibinfo{volume}{70}},
  \bibinfo{pages}{263} (\bibinfo{year}{1979}).

\bibitem[{\citenamefont{Pangali et~al.}(1979)\citenamefont{Pangali, Rao, and
  Berne}}]{Pangali:79}
\bibinfo{author}{\bibfnamefont{C.}~\bibnamefont{Pangali}},
  \bibinfo{author}{\bibfnamefont{M.}~\bibnamefont{Rao}}, \bibnamefont{and}
  \bibinfo{author}{\bibfnamefont{B.~J.} \bibnamefont{Berne}},
  \bibinfo{journal}{J. Chem. Phys.} \textbf{\bibinfo{volume}{71}},
  \bibinfo{pages}{2982} (\bibinfo{year}{1979}).

\bibitem[{\citenamefont{Zichi and Rossky}(1985)}]{Zichi:85}
\bibinfo{author}{\bibfnamefont{D.~A.} \bibnamefont{Zichi}} \bibnamefont{and}
  \bibinfo{author}{\bibfnamefont{P.~J.} \bibnamefont{Rossky}},
  \bibinfo{journal}{J. Chem. Phys.} \textbf{\bibinfo{volume}{83}},
  \bibinfo{pages}{797} (\bibinfo{year}{1985}).

\bibitem[{\citenamefont{Smith et~al.}(1992)\citenamefont{Smith, Zhang, and
  Haymet}}]{Smith.D:92}
\bibinfo{author}{\bibfnamefont{D.~E.} \bibnamefont{Smith}},
  \bibinfo{author}{\bibfnamefont{L.}~\bibnamefont{Zhang}}, \bibnamefont{and}
  \bibinfo{author}{\bibfnamefont{A.~D.~J.} \bibnamefont{Haymet}},
  \bibinfo{journal}{J. Am. Chem. Soc.} \textbf{\bibinfo{volume}{114}},
  \bibinfo{pages}{5875} (\bibinfo{year}{1992}).

\bibitem[{\citenamefont{Smith and Haymet}(1993)}]{Smith.D:93}
\bibinfo{author}{\bibfnamefont{D.~E.} \bibnamefont{Smith}} \bibnamefont{and}
  \bibinfo{author}{\bibfnamefont{A.~D.~J.} \bibnamefont{Haymet}},
  \bibinfo{journal}{J. Chem. Phys.} \textbf{\bibinfo{volume}{98}},
  \bibinfo{pages}{6445} (\bibinfo{year}{1993}).

\bibitem[{\citenamefont{Pearlman}(1993)}]{Pearlman:93}
\bibinfo{author}{\bibfnamefont{D.~A.} \bibnamefont{Pearlman}},
  \bibinfo{journal}{J. Chem. Phys.} \textbf{\bibinfo{volume}{98}},
  \bibinfo{pages}{8946} (\bibinfo{year}{1993}).

\bibitem[{\citenamefont{Forsman and J\"onsson}(1994)}]{Forsman:94}
\bibinfo{author}{\bibfnamefont{J.}~\bibnamefont{Forsman}} \bibnamefont{and}
  \bibinfo{author}{\bibfnamefont{B.}~\bibnamefont{J\"onsson}},
  \bibinfo{journal}{J. Chem. Phys.} \textbf{\bibinfo{volume}{101}},
  \bibinfo{pages}{5116} (\bibinfo{year}{1994}).

\bibitem[{\citenamefont{Skipper et~al.}(1996)\citenamefont{Skipper, Bridgeman,
  Buckingham, and Mancera}}]{Skipper:96}
\bibinfo{author}{\bibfnamefont{N.~T.} \bibnamefont{Skipper}},
  \bibinfo{author}{\bibfnamefont{C.~H.} \bibnamefont{Bridgeman}},
  \bibinfo{author}{\bibfnamefont{A.~D.} \bibnamefont{Buckingham}},
  \bibnamefont{and} \bibinfo{author}{\bibfnamefont{R.~L.}
  \bibnamefont{Mancera}}, \bibinfo{journal}{Faraday Discuss.}
  \textbf{\bibinfo{volume}{103}}, \bibinfo{pages}{141} (\bibinfo{year}{1996}).

\bibitem[{\citenamefont{L\"udemann et~al.}(1996)\citenamefont{L\"udemann,
  Schreiber, Abseher, and Steinhauser}}]{Luedemann:96}
\bibinfo{author}{\bibfnamefont{S.}~\bibnamefont{L\"udemann}},
  \bibinfo{author}{\bibfnamefont{H.}~\bibnamefont{Schreiber}},
  \bibinfo{author}{\bibfnamefont{R.}~\bibnamefont{Abseher}}, \bibnamefont{and}
  \bibinfo{author}{\bibfnamefont{O.}~\bibnamefont{Steinhauser}},
  \bibinfo{journal}{J. Chem. Phys.} \textbf{\bibinfo{volume}{104}},
  \bibinfo{pages}{286} (\bibinfo{year}{1996}).

\bibitem[{\citenamefont{L\"udemann et~al.}(1997)\citenamefont{L\"udemann,
  Abseher, Schreiber, and Steinhauser}}]{Luedemann:97}
\bibinfo{author}{\bibfnamefont{S.}~\bibnamefont{L\"udemann}},
  \bibinfo{author}{\bibfnamefont{R.}~\bibnamefont{Abseher}},
  \bibinfo{author}{\bibfnamefont{H.}~\bibnamefont{Schreiber}},
  \bibnamefont{and}
  \bibinfo{author}{\bibfnamefont{O.}~\bibnamefont{Steinhauser}},
  \bibinfo{journal}{J. Am. Chem. Soc.} \textbf{\bibinfo{volume}{119}},
  \bibinfo{pages}{4206} (\bibinfo{year}{1997}).

\bibitem[{\citenamefont{Ghosh et~al.}(2001)\citenamefont{Ghosh, Garc\'ia, and
  Garde}}]{Ghosh:2001}
\bibinfo{author}{\bibfnamefont{T.}~\bibnamefont{Ghosh}},
  \bibinfo{author}{\bibfnamefont{A.~E.} \bibnamefont{Garc\'ia}},
  \bibnamefont{and} \bibinfo{author}{\bibfnamefont{S.}~\bibnamefont{Garde}},
  \bibinfo{journal}{J. Am. Chem. Soc.} \textbf{\bibinfo{volume}{123}},
  \bibinfo{pages}{10997} (\bibinfo{year}{2001}).

\bibitem[{\citenamefont{Ghosh et~al.}(2002)\citenamefont{Ghosh, Garc\'ia, and
  Garde}}]{Ghosh:2002}
\bibinfo{author}{\bibfnamefont{T.}~\bibnamefont{Ghosh}},
  \bibinfo{author}{\bibfnamefont{A.~E.} \bibnamefont{Garc\'ia}},
  \bibnamefont{and} \bibinfo{author}{\bibfnamefont{S.}~\bibnamefont{Garde}},
  \bibinfo{journal}{J. Chem. Phys} \textbf{\bibinfo{volume}{116}},
  \bibinfo{pages}{2480} (\bibinfo{year}{2002}).

\bibitem[{\citenamefont{Ghosh et~al.}(2003)\citenamefont{Ghosh, Garc\'ia, and
  Garde}}]{Ghosh:2003}
\bibinfo{author}{\bibfnamefont{T.}~\bibnamefont{Ghosh}},
  \bibinfo{author}{\bibfnamefont{A.~E.} \bibnamefont{Garc\'ia}},
  \bibnamefont{and} \bibinfo{author}{\bibfnamefont{S.}~\bibnamefont{Garde}},
  \bibinfo{journal}{J. Phys. Chem. B} \textbf{\bibinfo{volume}{107}},
  \bibinfo{pages}{612} (\bibinfo{year}{2003}).

\bibitem[{\citenamefont{Shimizu and Chan}(2000)}]{Shimizu:2000}
\bibinfo{author}{\bibfnamefont{S.}~\bibnamefont{Shimizu}} \bibnamefont{and}
  \bibinfo{author}{\bibfnamefont{H.~S.} \bibnamefont{Chan}},
  \bibinfo{journal}{J. Chem. Phys.} \textbf{\bibinfo{volume}{113}},
  \bibinfo{pages}{4683} (\bibinfo{year}{2000}).

\bibitem[{\citenamefont{Shimizu and Chan}(2001)}]{Shimizu:2001}
\bibinfo{author}{\bibfnamefont{S.}~\bibnamefont{Shimizu}} \bibnamefont{and}
  \bibinfo{author}{\bibfnamefont{H.~S.} \bibnamefont{Chan}},
  \bibinfo{journal}{J. Am. Chem. Soc} \textbf{\bibinfo{volume}{123}},
  \bibinfo{pages}{2083} (\bibinfo{year}{2001}).

\bibitem[{\citenamefont{Shimizu and Chan}(2002)}]{Shimizu:2002}
\bibinfo{author}{\bibfnamefont{S.}~\bibnamefont{Shimizu}} \bibnamefont{and}
  \bibinfo{author}{\bibfnamefont{H.~S.} \bibnamefont{Chan}},
  \bibinfo{journal}{Proteins: Struct., Funct., Genet.}
  \textbf{\bibinfo{volume}{49}}, \bibinfo{pages}{560} (\bibinfo{year}{2002}).

\bibitem[{\citenamefont{Rick and Berne}(1997)}]{Rick:97}
\bibinfo{author}{\bibfnamefont{S.~W.} \bibnamefont{Rick}} \bibnamefont{and}
  \bibinfo{author}{\bibfnamefont{B.~J.} \bibnamefont{Berne}},
  \bibinfo{journal}{J. Phys. Chem. B} \textbf{\bibinfo{volume}{101}},
  \bibinfo{pages}{10488} (\bibinfo{year}{1997}).

\bibitem[{\citenamefont{Rick}(2000)}]{Rick:2000}
\bibinfo{author}{\bibfnamefont{S.~W.} \bibnamefont{Rick}}, \bibinfo{journal}{J.
  Phys. Chem. B} \textbf{\bibinfo{volume}{104}}, \bibinfo{pages}{6884}
  (\bibinfo{year}{2000}).

\bibitem[{\citenamefont{Rick}(2003)}]{Rick:2003}
\bibinfo{author}{\bibfnamefont{S.~W.} \bibnamefont{Rick}}, \bibinfo{journal}{J.
  Chem. Phys. B} \textbf{\bibinfo{volume}{107}}, \bibinfo{pages}{9853}
  (\bibinfo{year}{2003}).

\bibitem[{\citenamefont{Paschek}(2004{\natexlab{a}})}]{Paschek:2004:1}
\bibinfo{author}{\bibfnamefont{D.}~\bibnamefont{Paschek}}, \bibinfo{journal}{J.
  Chem. Phys.} \textbf{\bibinfo{volume}{120}}, \bibinfo{pages}{6674}
  (\bibinfo{year}{2004}{\natexlab{a}}).

\bibitem[{\citenamefont{Paschek}(2004{\natexlab{b}})}]{Paschek:2004:2}
\bibinfo{author}{\bibfnamefont{D.}~\bibnamefont{Paschek}}, \bibinfo{journal}{J.
  Chem. Phys.} \textbf{\bibinfo{volume}{120}}, \bibinfo{pages}{10605}
  (\bibinfo{year}{2004}{\natexlab{b}}).

\bibitem[{\citenamefont{Wilhelm et~al.}(1977)\citenamefont{Wilhelm, Battino,
  and Wilcox}}]{Wilhelm:77}
\bibinfo{author}{\bibfnamefont{E.}~\bibnamefont{Wilhelm}},
  \bibinfo{author}{\bibfnamefont{R.}~\bibnamefont{Battino}}, \bibnamefont{and}
  \bibinfo{author}{\bibfnamefont{R.~J.} \bibnamefont{Wilcox}},
  \bibinfo{journal}{Chem. Rev.} \textbf{\bibinfo{volume}{77}},
  \bibinfo{pages}{219} (\bibinfo{year}{1977}).

\bibitem[{\citenamefont{Rettich et~al.}(1981)\citenamefont{Rettich, Handa,
  Battino, and Wilhelm}}]{Rettich:81}
\bibinfo{author}{\bibfnamefont{T.~R.} \bibnamefont{Rettich}},
  \bibinfo{author}{\bibfnamefont{Y.}~\bibnamefont{Handa}},
  \bibinfo{author}{\bibfnamefont{R.}~\bibnamefont{Battino}}, \bibnamefont{and}
  \bibinfo{author}{\bibfnamefont{E.}~\bibnamefont{Wilhelm}},
  \bibinfo{journal}{J. Phys. Chem.} \textbf{\bibinfo{volume}{85}},
  \bibinfo{pages}{3230} (\bibinfo{year}{1981}).

\bibitem[{\citenamefont{Naghibi et~al.}(1986)\citenamefont{Naghibi, Dec, and
  Gill}}]{Naghibi:86}
\bibinfo{author}{\bibfnamefont{H.}~\bibnamefont{Naghibi}},
  \bibinfo{author}{\bibfnamefont{S.~F.} \bibnamefont{Dec}}, \bibnamefont{and}
  \bibinfo{author}{\bibfnamefont{S.~J.} \bibnamefont{Gill}},
  \bibinfo{journal}{J. Phys. Chem.} \textbf{\bibinfo{volume}{90}},
  \bibinfo{pages}{4621} (\bibinfo{year}{1986}).

\bibitem[{\citenamefont{Privalov}(1990)}]{Privalov:1990}
\bibinfo{author}{\bibfnamefont{P.~L.} \bibnamefont{Privalov}},
  \bibinfo{journal}{Crit. Rev. Biochem. Mol. Biol.}
  \textbf{\bibinfo{volume}{25}}, \bibinfo{pages}{281} (\bibinfo{year}{1990}).

\bibitem[{\citenamefont{Royer}(2002)}]{Royer:2002}
\bibinfo{author}{\bibfnamefont{C.~A.} \bibnamefont{Royer}},
  \bibinfo{journal}{Biochim. Biophys. Acta-Protein Struct. Molec. Enzym.}
  \textbf{\bibinfo{volume}{1595}}, \bibinfo{pages}{201} (\bibinfo{year}{2002}).

\bibitem[{\citenamefont{Smeller}(2002)}]{Smeller:2002}
\bibinfo{author}{\bibfnamefont{L.}~\bibnamefont{Smeller}},
  \bibinfo{journal}{Biochim. Biophys. Acta-Protein Struct. Molec. Enzym.}
  \textbf{\bibinfo{volume}{1595}}, \bibinfo{pages}{11} (\bibinfo{year}{2002}).

\bibitem[{\citenamefont{Ludwig}(1999)}]{Ludwig:1999}
\bibinfo{editor}{\bibfnamefont{H.}~\bibnamefont{Ludwig}}, ed.,
  \emph{\bibinfo{title}{Advances in High Pressure Bioscience and
  Biotechnology}} (\bibinfo{publisher}{Springer}, \bibinfo{address}{Heidelberg,
  Germany}, \bibinfo{year}{1999}).

\bibitem[{\citenamefont{Zipp and Kauzmann}(1973)}]{Zipp:1973}
\bibinfo{author}{\bibfnamefont{A.}~\bibnamefont{Zipp}} \bibnamefont{and}
  \bibinfo{author}{\bibfnamefont{W.}~\bibnamefont{Kauzmann}},
  \bibinfo{journal}{Biochemistry} \textbf{\bibinfo{volume}{12}},
  \bibinfo{pages}{4217} (\bibinfo{year}{1973}).

\bibitem[{\citenamefont{Hawley}(1971)}]{Hawley:71}
\bibinfo{author}{\bibfnamefont{S.~A.} \bibnamefont{Hawley}},
  \bibinfo{journal}{Biochemistry} \textbf{\bibinfo{volume}{10}},
  \bibinfo{pages}{2436} (\bibinfo{year}{1971}).

\bibitem[{\citenamefont{Brandts et~al.}(1970)\citenamefont{Brandts, Oliveira,
  and Westort}}]{Brandts:1970}
\bibinfo{author}{\bibfnamefont{J.~F.} \bibnamefont{Brandts}},
  \bibinfo{author}{\bibfnamefont{R.~J.} \bibnamefont{Oliveira}},
  \bibnamefont{and} \bibinfo{author}{\bibfnamefont{C.}~\bibnamefont{Westort}},
  \bibinfo{journal}{Biochemistry} \textbf{\bibinfo{volume}{9}},
  \bibinfo{pages}{1038} (\bibinfo{year}{1970}).

\bibitem[{\citenamefont{Silva et~al.}(2001)\citenamefont{Silva, Foguel, and
  Royer}}]{Silva:2001}
\bibinfo{author}{\bibfnamefont{J.~L.} \bibnamefont{Silva}},
  \bibinfo{author}{\bibfnamefont{D.}~\bibnamefont{Foguel}}, \bibnamefont{and}
  \bibinfo{author}{\bibfnamefont{C.~A.} \bibnamefont{Royer}},
  \bibinfo{journal}{Trends Biochem. Sci.} \textbf{\bibinfo{volume}{26}},
  \bibinfo{pages}{612} (\bibinfo{year}{2001}).

\bibitem[{\citenamefont{Silva et~al.}(1992)\citenamefont{Silva, Silveira,
  Correia, and Pontes}}]{Silva:1992}
\bibinfo{author}{\bibfnamefont{J.~L.} \bibnamefont{Silva}},
  \bibinfo{author}{\bibfnamefont{C.~F.} \bibnamefont{Silveira}},
  \bibinfo{author}{\bibfnamefont{A.}~\bibnamefont{Correia}}, \bibnamefont{and}
  \bibinfo{author}{\bibfnamefont{L.}~\bibnamefont{Pontes}},
  \bibinfo{journal}{J. Mol. Biol.} \textbf{\bibinfo{volume}{223}},
  \bibinfo{pages}{545} (\bibinfo{year}{1992}).

\bibitem[{\citenamefont{Peng et~al.}(1993)\citenamefont{Peng, Jonas, and
  Silva}}]{Peng:1993}
\bibinfo{author}{\bibfnamefont{X.}~\bibnamefont{Peng}},
  \bibinfo{author}{\bibfnamefont{J.}~\bibnamefont{Jonas}}, \bibnamefont{and}
  \bibinfo{author}{\bibfnamefont{J.~L.} \bibnamefont{Silva}},
  \bibinfo{journal}{Proc. Natl. Acad. Sci. USA} \textbf{\bibinfo{volume}{90}},
  \bibinfo{pages}{1776} (\bibinfo{year}{1993}).

\bibitem[{\citenamefont{Panick et~al.}(1999)\citenamefont{Panick, Vidugiris,
  Malessa, Rapp, Winter, and Royer}}]{Panick:1999}
\bibinfo{author}{\bibfnamefont{G.}~\bibnamefont{Panick}},
  \bibinfo{author}{\bibfnamefont{G.~J.~A.} \bibnamefont{Vidugiris}},
  \bibinfo{author}{\bibfnamefont{R.}~\bibnamefont{Malessa}},
  \bibinfo{author}{\bibfnamefont{G.}~\bibnamefont{Rapp}},
  \bibinfo{author}{\bibfnamefont{R.}~\bibnamefont{Winter}}, \bibnamefont{and}
  \bibinfo{author}{\bibfnamefont{C.~A.} \bibnamefont{Royer}},
  \bibinfo{journal}{Biochemistry} \textbf{\bibinfo{volume}{38}},
  \bibinfo{pages}{4157} (\bibinfo{year}{1999}).

\bibitem[{\citenamefont{Herberhold and Winter}(2002)}]{Herberhold:2002}
\bibinfo{author}{\bibfnamefont{H.}~\bibnamefont{Herberhold}} \bibnamefont{and}
  \bibinfo{author}{\bibfnamefont{R.}~\bibnamefont{Winter}},
  \bibinfo{journal}{Biochemistry} \textbf{\bibinfo{volume}{41}},
  \bibinfo{pages}{2396} (\bibinfo{year}{2002}).

\bibitem[{\citenamefont{Herberhold et~al.}(2004)\citenamefont{Herberhold,
  Royer, and Winter}}]{Herberhold:2004}
\bibinfo{author}{\bibfnamefont{H.}~\bibnamefont{Herberhold}},
  \bibinfo{author}{\bibfnamefont{C.~A.} \bibnamefont{Royer}}, \bibnamefont{and}
  \bibinfo{author}{\bibfnamefont{R.}~\bibnamefont{Winter}},
  \bibinfo{journal}{Biochemistry} \textbf{\bibinfo{volume}{43}},
  \bibinfo{pages}{3336} (\bibinfo{year}{2004}).

\bibitem[{\citenamefont{Ravindra et~al.}(2004)\citenamefont{Ravindra, Royer,
  and Winter}}]{Ravindra:2004}
\bibinfo{author}{\bibfnamefont{R.}~\bibnamefont{Ravindra}},
  \bibinfo{author}{\bibfnamefont{C.}~\bibnamefont{Royer}}, \bibnamefont{and}
  \bibinfo{author}{\bibfnamefont{R.}~\bibnamefont{Winter}},
  \bibinfo{journal}{Phys. Chem. Chem. Phys.} \textbf{\bibinfo{volume}{6}},
  \bibinfo{pages}{1952} (\bibinfo{year}{2004}).

\bibitem[{\citenamefont{Hummer et~al.}(1998)\citenamefont{Hummer, Garde,
  Garc\'ia, Paulaitis, and Pratt}}]{Hummer:98:1}
\bibinfo{author}{\bibfnamefont{G.}~\bibnamefont{Hummer}},
  \bibinfo{author}{\bibfnamefont{S.}~\bibnamefont{Garde}},
  \bibinfo{author}{\bibfnamefont{A.~E.} \bibnamefont{Garc\'ia}},
  \bibinfo{author}{\bibfnamefont{M.~E.} \bibnamefont{Paulaitis}},
  \bibnamefont{and} \bibinfo{author}{\bibfnamefont{L.~R.} \bibnamefont{Pratt}},
  \bibinfo{journal}{Proc. Natl. Acad. Sci. USA} \textbf{\bibinfo{volume}{95}},
  \bibinfo{pages}{1552} (\bibinfo{year}{1998}).

\bibitem[{\citenamefont{Huang and Chandler}(2000)}]{Huang:2000}
\bibinfo{author}{\bibfnamefont{D.~M.} \bibnamefont{Huang}} \bibnamefont{and}
  \bibinfo{author}{\bibfnamefont{D.}~\bibnamefont{Chandler}},
  \bibinfo{journal}{Proc. Natl. Acad. Sci. USA} \textbf{\bibinfo{volume}{97}},
  \bibinfo{pages}{8324} (\bibinfo{year}{2000}).

\bibitem[{\citenamefont{{Rein ten Wolde} and
  Chandler}(2002)}]{Rein_ten_Wolde:2002}
\bibinfo{author}{\bibfnamefont{P.}~\bibnamefont{{Rein ten Wolde}}}
  \bibnamefont{and} \bibinfo{author}{\bibfnamefont{D.}~\bibnamefont{Chandler}},
  \bibinfo{journal}{Proc. Natl. Acad. Sci. USA} \textbf{\bibinfo{volume}{99}},
  \bibinfo{pages}{6539} (\bibinfo{year}{2002}).

\bibitem[{\citenamefont{Chandler}(2005)}]{Chandler:2005}
\bibinfo{author}{\bibfnamefont{D.}~\bibnamefont{Chandler}},
  \bibinfo{journal}{Nature}  (\bibinfo{year}{2005}), \bibinfo{note}{submitted
  insight review article}.

\bibitem[{\citenamefont{Ghosh et~al.}(2005)\citenamefont{Ghosh, Kalra, and
  Garde}}]{Ghosh:2005}
\bibinfo{author}{\bibfnamefont{T.}~\bibnamefont{Ghosh}},
  \bibinfo{author}{\bibfnamefont{A.}~\bibnamefont{Kalra}}, \bibnamefont{and}
  \bibinfo{author}{\bibfnamefont{S.}~\bibnamefont{Garde}}, \bibinfo{journal}{J.
  Phys. Chem. B} \textbf{\bibinfo{volume}{109}}, \bibinfo{pages}{642}
  (\bibinfo{year}{2005}).

\bibitem[{\citenamefont{Mahoney and Jorgensen}(2000)}]{Mahoney:2000}
\bibinfo{author}{\bibfnamefont{M.~W.} \bibnamefont{Mahoney}} \bibnamefont{and}
  \bibinfo{author}{\bibfnamefont{W.~L.} \bibnamefont{Jorgensen}},
  \bibinfo{journal}{J. Chem. Phys.} \textbf{\bibinfo{volume}{112}},
  \bibinfo{pages}{8910} (\bibinfo{year}{2000}).

\bibitem[{\citenamefont{Nos\'e}(1984)}]{Nose:84}
\bibinfo{author}{\bibfnamefont{S.}~\bibnamefont{Nos\'e}},
  \bibinfo{journal}{Mol. Phys.} \textbf{\bibinfo{volume}{52}},
  \bibinfo{pages}{255} (\bibinfo{year}{1984}).

\bibitem[{\citenamefont{Hoover}(1985)}]{Hoover:85}
\bibinfo{author}{\bibfnamefont{W.~G.} \bibnamefont{Hoover}},
  \bibinfo{journal}{Phys. Rev. A} \textbf{\bibinfo{volume}{31}},
  \bibinfo{pages}{1695} (\bibinfo{year}{1985}).

\bibitem[{\citenamefont{Parrinello and Rahman}(1981)}]{Parrinello:81}
\bibinfo{author}{\bibfnamefont{M.}~\bibnamefont{Parrinello}} \bibnamefont{and}
  \bibinfo{author}{\bibfnamefont{A.}~\bibnamefont{Rahman}},
  \bibinfo{journal}{J. Appl. Phys.} \textbf{\bibinfo{volume}{52}},
  \bibinfo{pages}{7182} (\bibinfo{year}{1981}).

\bibitem[{\citenamefont{Nos\'e and Klein}(1983)}]{Nose:83}
\bibinfo{author}{\bibfnamefont{S.}~\bibnamefont{Nos\'e}} \bibnamefont{and}
  \bibinfo{author}{\bibfnamefont{M.~L.} \bibnamefont{Klein}},
  \bibinfo{journal}{Mol. Phys.} \textbf{\bibinfo{volume}{50}},
  \bibinfo{pages}{1055} (\bibinfo{year}{1983}).

\bibitem[{\citenamefont{Essmann et~al.}(1995)\citenamefont{Essmann, Perera,
  Berkowitz, Darden, Lee, and Pedersen}}]{Essmann:95}
\bibinfo{author}{\bibfnamefont{U.}~\bibnamefont{Essmann}},
  \bibinfo{author}{\bibfnamefont{L.}~\bibnamefont{Perera}},
  \bibinfo{author}{\bibfnamefont{M.~L.} \bibnamefont{Berkowitz}},
  \bibinfo{author}{\bibfnamefont{T.~A.} \bibnamefont{Darden}},
  \bibinfo{author}{\bibfnamefont{H.}~\bibnamefont{Lee}}, \bibnamefont{and}
  \bibinfo{author}{\bibfnamefont{L.~G.} \bibnamefont{Pedersen}},
  \bibinfo{journal}{J. Chem. Phys.} \textbf{\bibinfo{volume}{103}},
  \bibinfo{pages}{8577} (\bibinfo{year}{1995}).

\bibitem[{\citenamefont{Miyamoto and Kollman}(1992)}]{Miyamoto:92}
\bibinfo{author}{\bibfnamefont{S.}~\bibnamefont{Miyamoto}} \bibnamefont{and}
  \bibinfo{author}{\bibfnamefont{P.~A.} \bibnamefont{Kollman}},
  \bibinfo{journal}{J. Comp. Chem.} \textbf{\bibinfo{volume}{13}},
  \bibinfo{pages}{952} (\bibinfo{year}{1992}).

\bibitem[{\citenamefont{Ryckaert et~al.}(1977)\citenamefont{Ryckaert, Ciccotti,
  and Berendsen}}]{Ryckaert:77}
\bibinfo{author}{\bibfnamefont{J.~P.} \bibnamefont{Ryckaert}},
  \bibinfo{author}{\bibfnamefont{G.}~\bibnamefont{Ciccotti}}, \bibnamefont{and}
  \bibinfo{author}{\bibfnamefont{H.~J.~C.} \bibnamefont{Berendsen}},
  \bibinfo{journal}{J. Comp. Phys.} \textbf{\bibinfo{volume}{23}},
  \bibinfo{pages}{327} (\bibinfo{year}{1977}).

\bibitem[{\citenamefont{Lindahl et~al.}(2001)\citenamefont{Lindahl, Hess, and
  {van der Spoel}}}]{gmxpaper}
\bibinfo{author}{\bibfnamefont{E.}~\bibnamefont{Lindahl}},
  \bibinfo{author}{\bibfnamefont{B.}~\bibnamefont{Hess}}, \bibnamefont{and}
  \bibinfo{author}{\bibfnamefont{D.}~\bibnamefont{{van der Spoel}}},
  \bibinfo{journal}{J. Mol. Model.} \textbf{\bibinfo{volume}{7}},
  \bibinfo{pages}{306} (\bibinfo{year}{2001}).

\bibitem[{\citenamefont{{van der Spoel} et~al.}(2004)\citenamefont{{van der
  Spoel}, Lindahl, Hess, {van Buuren}, Apol, Meulenhoff, Tieleman, Sijbers,
  Feenstra, {van Drunen} et~al.}}]{gmx32}
\bibinfo{author}{\bibfnamefont{D.}~\bibnamefont{{van der Spoel}}},
  \bibinfo{author}{\bibfnamefont{E.}~\bibnamefont{Lindahl}},
  \bibinfo{author}{\bibfnamefont{B.}~\bibnamefont{Hess}},
  \bibinfo{author}{\bibfnamefont{A.~R.} \bibnamefont{{van Buuren}}},
  \bibinfo{author}{\bibfnamefont{E.}~\bibnamefont{Apol}},
  \bibinfo{author}{\bibfnamefont{P.~J.} \bibnamefont{Meulenhoff}},
  \bibinfo{author}{\bibfnamefont{D.~P.} \bibnamefont{Tieleman}},
  \bibinfo{author}{\bibfnamefont{A.~L. T.~M.} \bibnamefont{Sijbers}},
  \bibinfo{author}{\bibfnamefont{K.~A.} \bibnamefont{Feenstra}},
  \bibinfo{author}{\bibfnamefont{R.}~\bibnamefont{{van Drunen}}},
  \bibnamefont{et~al.}, \emph{\bibinfo{title}{Gromacs {U}ser {M}anual version
  3.2}}, \bibinfo{address}{www.gromacs.org} (\bibinfo{year}{2004}).

\bibitem[{\citenamefont{Flyvbjerg and Petersen}(1989)}]{Flyvbjerg:89}
\bibinfo{author}{\bibfnamefont{H.}~\bibnamefont{Flyvbjerg}} \bibnamefont{and}
  \bibinfo{author}{\bibfnamefont{H.~G.} \bibnamefont{Petersen}},
  \bibinfo{journal}{J. Chem. Phys.} \textbf{\bibinfo{volume}{91}},
  \bibinfo{pages}{461} (\bibinfo{year}{1989}).

\bibitem[{\citenamefont{Berendsen et~al.}(1984)\citenamefont{Berendsen, Postma,
  van Gunsteren, DiNola, and Haak}}]{Berendsen:84}
\bibinfo{author}{\bibfnamefont{H.~J.~C.} \bibnamefont{Berendsen}},
  \bibinfo{author}{\bibfnamefont{J.~P.~M.} \bibnamefont{Postma}},
  \bibinfo{author}{\bibfnamefont{W.~F.} \bibnamefont{van Gunsteren}},
  \bibinfo{author}{\bibfnamefont{A.}~\bibnamefont{DiNola}}, \bibnamefont{and}
  \bibinfo{author}{\bibfnamefont{J.~R.} \bibnamefont{Haak}},
  \bibinfo{journal}{J. Chem. Phys.} \textbf{\bibinfo{volume}{81}},
  \bibinfo{pages}{3684} (\bibinfo{year}{1984}).

\bibitem[{\citenamefont{Neumann}(1983)}]{Neumann:83}
\bibinfo{author}{\bibfnamefont{M.}~\bibnamefont{Neumann}},
  \bibinfo{journal}{Mol. Phys.} \textbf{\bibinfo{volume}{50}},
  \bibinfo{pages}{841} (\bibinfo{year}{1983}).

\bibitem[{\citenamefont{Roberts and Schnitker}(1994)}]{Roberts:94}
\bibinfo{author}{\bibfnamefont{J.~E.} \bibnamefont{Roberts}} \bibnamefont{and}
  \bibinfo{author}{\bibfnamefont{J.}~\bibnamefont{Schnitker}},
  \bibinfo{journal}{J. Chem. Phys.} \textbf{\bibinfo{volume}{101}},
  \bibinfo{pages}{5024} (\bibinfo{year}{1994}).

\bibitem[{\citenamefont{Roberts and Schnitker}(1995)}]{Roberts:95}
\bibinfo{author}{\bibfnamefont{J.~E.} \bibnamefont{Roberts}} \bibnamefont{and}
  \bibinfo{author}{\bibfnamefont{J.}~\bibnamefont{Schnitker}},
  \bibinfo{journal}{J. Phys. Chem.} \textbf{\bibinfo{volume}{99}},
  \bibinfo{pages}{1322} (\bibinfo{year}{1995}).

\bibitem[{\citenamefont{Press et~al.}(1992)\citenamefont{Press, Teukolsky,
  Vetterling, and Flannery}}]{NumRecipes}
\bibinfo{author}{\bibfnamefont{W.~H.} \bibnamefont{Press}},
  \bibinfo{author}{\bibfnamefont{S.~A.} \bibnamefont{Teukolsky}},
  \bibinfo{author}{\bibfnamefont{W.~T.} \bibnamefont{Vetterling}},
  \bibnamefont{and} \bibinfo{author}{\bibfnamefont{B.~P.}
  \bibnamefont{Flannery}}, \emph{\bibinfo{title}{Numerical Recipies: The art of
  scientific computing}} (\bibinfo{publisher}{Cambridge University Press},
  \bibinfo{address}{Cambridge}, \bibinfo{year}{1992}), \bibinfo{edition}{2nd}
  ed.

\bibitem[{\citenamefont{Nicolini et~al.}(2004)\citenamefont{Nicolini, Ravindra,
  Ludolph, and Winter}}]{Nicolini:2004}
\bibinfo{author}{\bibfnamefont{C.}~\bibnamefont{Nicolini}},
  \bibinfo{author}{\bibfnamefont{R.}~\bibnamefont{Ravindra}},
  \bibinfo{author}{\bibfnamefont{B.}~\bibnamefont{Ludolph}}, \bibnamefont{and}
  \bibinfo{author}{\bibfnamefont{R.}~\bibnamefont{Winter}},
  \bibinfo{journal}{Biophys. J.} \textbf{\bibinfo{volume}{86}},
  \bibinfo{pages}{1385} (\bibinfo{year}{2004}).

\bibitem[{\citenamefont{Daggett and Levitt}(1992)}]{Daggett:1992}
\bibinfo{author}{\bibfnamefont{V.}~\bibnamefont{Daggett}} \bibnamefont{and}
  \bibinfo{author}{\bibfnamefont{M.}~\bibnamefont{Levitt}},
  \bibinfo{journal}{Proc. Natl. Acad. Sci. USA} \textbf{\bibinfo{volume}{89}},
  \bibinfo{pages}{5142} (\bibinfo{year}{1992}).

\bibitem[{\citenamefont{Urry}(1993)}]{Urry:93}
\bibinfo{author}{\bibfnamefont{D.~W.} \bibnamefont{Urry}},
  \bibinfo{journal}{Angew. Chem.-Int. Edit. Engl.}
  \textbf{\bibinfo{volume}{32}}, \bibinfo{pages}{819} (\bibinfo{year}{1993}).

\bibitem[{\citenamefont{Silverstein et~al.}(2000)\citenamefont{Silverstein,
  Haymet, and Dill}}]{Silverstein:2000}
\bibinfo{author}{\bibfnamefont{K.~A.~T.} \bibnamefont{Silverstein}},
  \bibinfo{author}{\bibfnamefont{A.~D.~J.} \bibnamefont{Haymet}},
  \bibnamefont{and} \bibinfo{author}{\bibfnamefont{K.~A.} \bibnamefont{Dill}},
  \bibinfo{journal}{J. Am. Chem. Soc.} \textbf{\bibinfo{volume}{122}},
  \bibinfo{pages}{8037} (\bibinfo{year}{2000}).

\end{thebibliography}
\end{document}